    \documentstyle[12pt, aaspp4]  {article}

\begin{document}

\title{First Astronomical Detection of the Cumulene Carbon Chain Molecule
H$_2$C$_6$ in TMC1}
\author{W. D. Langer\altaffilmark{1}, T. Velusamy, T. B. H. Kuiper, and R. Peng}
\affil{Jet Propulsion Laboratory, California Institute of Technology,
MS 169-506, Pasadena, CA 91109}
\and
\author{M. C. McCarthy\altaffilmark{2}, M. J. Travers\altaffilmark{2},
A. Kov\'{a}cs, C. A. Gottlieb, and P. Thaddeus\altaffilmark{2}}
\affil{Division of Engineering and Applied Sciences, Harvard University, 29 Oxford
Street, Cambridge, MA 02138}
\altaffiltext{1}{langer@langer.jpl.nasa.gov}
\altaffiltext{2}{Harvard-Smithsonian Center for Astrophysics, 60 Garden Street,
Cambridge, MA 02138}

\begin{abstract}

The cumulene carbenes are important components of hydrocarbon chemistry in low mass
star forming cores.  Here we report the first astronomical detection of the long chain
cumulene carbene H$_2$C$_6$ in the interstellar cloud TMC1, from observations of two
of its rotational transitions: $J_{K,K'} = 7_{1,7} \rightarrow 6_{1,6}$ at 18.8 GHz
and $8_{1,8} \rightarrow 7_{1,7}$ at 21.5 GHz, using NASA's Deep Space Network 70 m
antenna at Goldstone, California.  In addition we also observed the shorter cumulene
carbene, H$_2$C$_4$ at the same position.  The fractional abundance of H$_2$C$_6$
relative to H$_2$ is about $4.7 \times 10^{-11}$ and H$_2$C$_4$ is about $1.1
\times 10^{-9}$.  The abundance of H$_2$C$_6$ is in fairly good agreement with gas
phase chemical models for young molecular cloud cores, but the abundance of
H$_2$C$_4$ is significantly larger than predicted.

\end{abstract}


\lefthead{Langer et al.}
\righthead{First Astronomical Detection of H$_2$C$_6$}

\pagebreak

\section{INTRODUCTION}

The detection of new cumulene carbenes is important for determining the
hydrocarbon chemistry in low mass star forming cores (cf.  Bettens \& Herbst
1996).	The cumulene carbene chains are also of interest in astronomy for their
possible role as carriers of the diffuse interstellar bands (McCarthy {\em et
al.} 1997).  Here we report the first astronomical detection of H$_2$C$_6$
(hexapentaenylidene) in the dark cloud TMC1.  The identification of this highly
reactive isomer of triacetylene (HC$_6$H) was made possible by the recent
laboratory detection of the rotational spectrum of both H$_2$C$_5$ and
H$_2$C$_6$ (McCarthy et al.  1997), and the high sensitivity of NASA's Deep
Space Network antennas at cm wavelengths.

The shorter cumulene carbenes propadienylidene H$_2$C$_3$ (Cernicharo {\em et al.}
1991, Kawaguchi {\em et al.} 1991) and butatrienylidene H$_2$C$_4$ (Kawaguchi {\em et
al.}) were also observed at the same position as H$_2$C$_6$, allowing
comparison with abundances predicted by chemical models.  Both H$_2$C$_4$ and
H$_2$C$_6$ have linear carbon backbones and C$_{2v}$ symmetry and, because of the
off-axis equivalent H-atoms, they both have ortho ($K = \pm 1$, odd) and para
($K = 0$, even) rotational ladders between which both radiative and collisional
transitions are forbidden.  In cold molecular clouds we expect 3$/$4 of the population
to be in the lowest lying ortho ($K = \pm 1$) ladders, and emission from these to be
slightly stronger than from the para ($K = 0$) ladder.

In addition to H$_2$C$_6$, several other complex molecules were observed in
TMC1, for the purpose of clarifying hydrocarbon chemistry in these star
forming cores.  We find that the carbon chain molecules vary in abundance among the
three principal velocity components.  The distribution of the cumulene carbenes 
is similar to those of other complex carbon compounds, such as the
cyanopolyynes HC$_{2n+1}$N and the carbon chain radicals C$_n$H, but differ
from the cyclopropenylidene ring, $c$-C$_3$H$_2$.  The observed ratios of
abundances in TMC1 are compared with gas phase chemical models.

\section{OBSERVATIONS}

The observations were made with NASA's Deep Space Network (DSN) 70 meter antenna at
Goldstone, CA during October and November of 1996.	In addition to H$_2$C$_6$,
five other molecules (HC$_9$N, C$_5$H, H$_2$C$_3$, H$_2$C$_4$, and $c$-C$_3$H$_2$)
were observed as part of an ongoing effort to study the carbon chemistry
in TMC1.  The relevant transitions and line frequencies are listed in
Table~1.

The observations were made with a broad band (17.5 - 26 GHz) cooled HEMT receiver with
typical system temperatures of 50 - 65 K and the 2 million channel Wide Band
Spectrometer Analyzer (WBSA).  The spectra were smoothed to resolutions in the range
of 0.03 to 0.15 km s$^{-1}$ with a total spectral range of 20 to 40 MHz (cf.  Langer
{\em et al.} 1995).	All observations, except that of the 18.4 GHz transition of
H$_2$C$_6$, were made using position switching.  In this mode the spectra were
obtained by taking the
 difference between successive ON and OFF source integrations. To suppress
baseline irregularities of instrumental and atmospheric origin, the right ascension of
the OFF source position was also chosen to be 6 minutes in time to the west or east of
the ON source position such that each ON-OFF pair had the same mean elevation.  The
presence of several stronger lines in the passband of the 21.5 GHz line of H$_2$C$_6$
and the 20.8 GHz line of H$_2$C$_3$, allowed us to check our calibration
during each observing session. Following the detection of the 21.5 GHz
line of H$_2$C$_6$, a frequency switching option was implemented for the DSN-WBSA
system.  We used this mode for the observations of the 18.8 GHz transition of
H$_2$C$_6$ using an LO offset of 0.5 MHz between the signal and reference
frequencies.

The detections reported here were made toward TMC1 at a position [$\alpha$(1950) =
4$^h$38$^m$41$^s$ and $\delta$(1950) = 25$^o$35$\arcmin$40$\arcsec$] corresponding to
the strongest HC$_7$N and HC$_9$N emission observed at 
$v_{\rm{LSR}} = 5.8$~km~s$^{-1}$ in our DSN 70~m high spectral resolution maps
(Velusamy {\em et al.} 1997). The ON source integration times for the H$_2$C$_6$ lines
were 23 and 18 hours for the 21.5 and 18.8 GHz transitions, respectively, and yielded
a corresponding rms of 2.5 mK and 3.5 mK at spectral resolutions of 0.14 and
0.12 km s$^{-1}$.

\section{RESULTS}

In Figure 1a we show the spectra for the two transitions of H$_2$C$_6$, and those
of C$_5$H and HC$_9$N.  The H$_2$C$_6$ 21.5 GHz transition, HC$_9$N, and C$_5$H were
observed simultaneously in the same passband.  The spectrum of the $2_{0,2}
\rightarrow 1_{0,1}$ transition of H$_2$C$_4$  at 17.9 GHz in TMC1
is shown in Figure~1b, along with the lines of H$_2$C$_3$, C$_6$H, and
$\it{c}$-C$_3$H$_2$.  The H$_2$C$_4$ and H$_2$C$_6$ spectra peak at $\sim 6$ km
s$^{-1}$ with antenna temperatures T$_A = 180$ mK and 9 mK, respectively.  Previous
high spectral resolution studies of carbon chain molecules in TMC1, such
as CCS and HC$_7$N (Langer {\em et al.} 1995), revealed that there are three
velocity components at 5.7, 5.9, and 6.1 km s$^{-1}$ with very narrow linewidths
$\sim 0.15$ to 0.20 km s$^{-1}$.  Our detection of H$_2$C$_6$ in TMC1 supports the
suggestion of a tentative assignment of a weak unblended and a blended 3~mm line in
IRC+10216 (Gu\'{e}lin et al.  1997).

In Figure~1 it appears that the strongest H$_2$C$_4$ and H$_2$C$_6$ emission is
at 5.9 km s$^{-1}$.  The slight asymmetry in the line profiles is
most likely due to a second weaker (partially blended) component at 6.1
km s$^{-1}$, exactly what is seen in the velocity structure of C$_5$H and
probably C$_6$H.  There is no evidence in H$_2$C$_4$ and H$_2$C$_6$ for a
component at 5.7 km s$^{-1}$.  In contrast, $\it{c}$-C$_3$H$_2$ and its
linear isomer H$_2$C$_3$ show strong emission at 5.7 and 5.9 km s$^{-1}$,
and only slightly less emission at 6.1 km s$^{-1}$.  The presence of strong
$\it{c}$-C$_3$H$_2$ and H$_2$C$_3$ emission at 5.7 km s$^{-1}$ and the
apparent absence of H$_2$C$_6$ or H$_2$C$_4$ emission at that velocity
indicates that there is a significant chemical inhomogeneity in the hydrocarbon
chemistry in TMC1.  Furthermore, the $\it{c}$-C$_3$H$_2$ emission is much
stronger than that of H$_2$C$_3$ (T$_A = 2.0$ K versus 0.12 K).

\section{DISCUSSION}

To estimate the abundances of H$_2$C$_6$ and H$_2$C$_4$ in TMC1 we adopted a
simple model in which we treat each rotational ladder as a separate
linear rotor.  The excitation of the ortho and para states can be treated
separately, because the radiative and collisional transitions between them are
negligible.  For purposes of estimating abundances, the excitation of the ortho
ladder can be treated as two equivalent linear rotors.	The energy levels of
the $K = \pm 1$ ladders are nearly equal and can be estimated within 1\% by
treating them as a linear rotor with $B_0 = (B + C)/2$ (see McCarthy {\em et
al.}).	Collisional rate coefficients for these species have not been
calculated; we approximate their $J$-dependent values by using those derived
for the linear chain molecule HC$_3$N (Green and Chapman 1978).  We assume that
the rotational population of H$_2$C$_6$ has a normal ortho:para ratio of 3:1.
Furthermore, reactive collisions with protonating ions can produce
H$_2$C$_6$H$^+$ which, upon decomposition to H$_2$C$_6$ + H, will preserve the
nuclear spins (Herbst, 1977).

H$_2$C$_6$ is highly polar, with a calculated dipole moment of 6.2~D (Maluendes
\& McLean 1992).  The $J = 8$ ortho levels lie $\sim 5$ K above the ortho ground state
($J = 1$) and these levels have an Einstein~$A$ coefficient of about $2 \times
10^{-6}$ s$^{-1}$.  This yields a critical density $n$(H$_2) \sim 5 \times
10^3$~cm$^{-3}$, implying that the lower rotational levels are easily thermalized at
the density in the cores of TMC1. Figure 2 shows an excitation calculation for the $K
= -1$ ortho ladder of H$_2$C$_6$ and the $K = 0$ para ladder of H$_2$C$_4$ (using a
dipole moment of 4.5~D) for a range of fractional abundance gradients (in km s$^{-1}$
pc$^{-1}$).  For convenience, calculations were done with a large velocity
gradient (LVG) code which is essentially equivalent to an LTE calculation for
optically thin lines, as appropriate for the weak emission seen in H$_2$C$_4$ and
H$_2$C$_6$.  We assumed typical conditions found in the TMC1 clumps:  T$_{kin} = 10$ K
and $n$(H$_2) = 10^4$~cm$^{-3}$ (Langer et al.).  The velocity gradient is estimated
from the line width, $\sim 0.3$ km s$^{-1}$, divided by the size of the emission
region ($\sim 0.06$~pc) estimated from our HC$_7$N map.  The total fractional
abundances of H$_2$C$_4$ and H$_2$C$_6$, after accounting for molecules in the other
ladders, a velocity gradient of 5 km s$^{-1}$ pc$^{-1}$, and a beam efficiency of 0.7,
are about $1.1 \times 10^{-9}$ and $4.7 \times 10^{-11}$, respectively.  The column
density for H$_2$C$_4$ derived by Kawaguchi et al.	(1991) from several transitions at a
nearby position corresponds to a fractional abundance of $8 \times 10^{-10}$ (Bettens
et al.  1995), which is in very good agreement with our value. We estimate that
H$_2$C$_4$ is $\sim 25$ more abundant than H$_2$C$_6$.

Recent gas phase chemical models have included reaction networks for
calculating abundances of long carbon chains (Millar {\em et al.} 1997,
Bettens, Lee \& Herbst 1995) and very large hydrocarbons (Thaddeus 1994, and
Bettens \& Herbst 1996).  Furthermore, laboratory data and theoretical
calculations have begun to clarify the mechanisms for forming the simplest ring
structure cyclopropynylidyne, $\it{c}$-C$_3$H, via neutral carbon insertion
into acetylene, C + C$_2$H$_2$ (Kaiser {\em et al.} 1996).  To compare
qualitatively our observations with chemical models we use the results of the
gas phase model of Millar et al.  (1997), which is based on one of the standard
chemical data bases used for modeling cloud chemistry.	It includes hydrocarbon
production up to 8 or 9 carbons, but does not include the carbon insertion
reactions.

In Figure~3 we plot these model abundances for the carbon chains C$_{2n}$H,
C$_{2n+1}$H, H$_2$C$_{2n}$, H$_2$C$_{2n+1}$, and HC$_{2n+1}$N, as well as a point for
$c$-C$_3$H$_2$, for the physical conditions relevant to this core of TMC1.  Figure~3
uses the early time results of Millar et al.  as only this evolutionary stage has
enough neutral carbon to produce a successful complex carbon chemistry with large
hydrocarbon abundances.  The late times, or steady state, solutions have too little
carbon and complex carbon species to explain the observations.  This result is not
surprising as other tracers, such as CCS, confirm the ages of these dense core
components to be $\leq$ a few
$\times 10^5$ years (Kuiper {\em et al.} 1996, Velusamy {\em et al.} 1997). Note that
the C$_n$H radicals are predicted to be more abundant than the corresponding
H$_2$C$_n$ cumulene carbenes, as observed.  However, the Millar et al. model
also predicts that H$_2$C$_6$ is more abundant than H$_2$C$_4$ which is exactly the
opposite from what is observed.  The measured H$_2$C$_6$ abundance agrees within
a factor of 2 with their model calculations, while the predicted H$_2$C$_4$ fractional
abundance is too low by a factor $\sim$ 25.  On the other hand, the observed
H$_2$C$_4/$H$_2$C$_6$ ratio is consistent with the slope of the abundance
versus chain length in the C$_n$H radicals. However, some caution is in order
regarding absolute abundances, because this model does not distinguish between
different isomers, such as triacetylene (HC$_6$H) and H$_2$C$_6$.

Bettens et al.	explored the effects of various assumptions about the
neutral-neutral reactions on the production of complex carbon molecules in
interstellar clouds.  Bettens \& Herbst (1996) recommend using two of their
chemical models which best explain the polyatomic species:  the new standard
model (NSM) and model~4 (M4).  NSM is a modified version of the standard
ion-molecule chemical scheme; in M4, rapid neutral-neutral reactions play a
critical role.	A key feature of M4 is that the reactions O + C$_n$
$\rightarrow$ C$_{n-1}$ + CO are assumed negligible for $n > 2$.  Model M4 of
Bettens et al.	provides the best agreement with our H$_2$C$_4$ observations,
but the calculated abundance is still too high by a factor of five.  The
models appear to be sensitive to assumptions about the neutral-neutral reaction
rate coefficients at low temperature.  H$_2$C$_6$ and H$_2$C$_4$ are important
diagnostics to discriminate among models of hydrocarbon chemistry in dense
cores, however, more laboratory measurements and astronomical observations are
needed to resolve the hydrocarbon chemistry in dense cores.

We thank the staff of the DSN at Goldstone for their assistance in making these
observations and Eric Herbst for useful comments.  The research of the JPL
group was conducted at the Jet Propulsion Laboratory, California Institute of
Technology, under contract with the National Aeronautics and Space
Administration.  WDL would also like to thank the Smithsonian Astrophysical
Observatory and the Center for Astrophysics for their hospitality during two
visits while this work was in progress.

\clearpage

\begin{table*}
\centering
\begin{tabular}{llll}
\multicolumn{4}{c}{TABLE 1} \\ \\
\multicolumn{4}{c}{MOLECULAR TRANSITIONS OBSERVED IN TMC1} \\ \\
\hline
\multicolumn{1}{c}{Species} &
\multicolumn{1}{c}{Transition} &
\multicolumn{1}{c}{Frequency} &
\multicolumn{1}{c}{Reference} \\
\multicolumn{1}{c}{} &
\multicolumn{1}{c}{} &
\multicolumn{1}{c}{MHz} &
\multicolumn{1}{c}{} \\
\tableline
H$_2$C$_3$     & $1_{0,1} \rightarrow 0_{0,0}$       & 20792.640 &Lovas et al. \\
$c$-C$_3$H$_2$ & $1_{1,0} \rightarrow 1_{0,1}$       & 18343.143 &Pickett et al. \\
H$_2$C$_4$     & $2_{0,2} \rightarrow 1_{0,1}$       & 17863.810 &Travers et al. \\
C$_5$H         & $J = 9/2 \rightarrow 7/2, F=5-4b$   & 21484.710 &Lovas \\
C$_5$H         & $J = 7/2 \rightarrow 7/2, F=4-3b$   & 21485.262 &Lovas \\
C$_6$H         & $J = 15/2 \rightarrow 13/2, F=8-7f$ & 20792.872 &Lovas \\
H$_2$C$_6$     & $7_{1,7} \rightarrow 6_{1,6}$       & 18802.235 &McCarthy et al. \\
H$_2$C$_6$     & $8_{1,8} \rightarrow 7_{1,7}$       & 21488.255 &McCarthy et al. \\
HC$_9$N        & $37 \rightarrow 36$                 & 21498.181 & Pickett et al. \\
\hline
\end{tabular}
\end{table*}

\clearpage

\begin{figure}[tbp]
\caption{(a) DSN 70 m detections of H$_2$C$_6$ at 21.5 and 18.8 GHz in TMC1. The
spectra for H$_2$C$_6$ at 21.5 GHz, HC$_9$N, and C$_5$H, were taken within the same
passband, while the 18.8 GHz line of H$_2$C$_6$ was observed separately.  The source
position is $\alpha$(1950) = 4$^h$38$^m$41$^s$ and
$\delta$(1950) = 25$^o $35$\arcmin$40$\arcsec$, with $v_{{\rm LSR}} = 5.8$~km~s$^{-1}$.
(b) The spectrum for H$_2$C$_4$ towards TMC1 is shown along with those of C$_6$H (2
hyperfine components), H$_2$C$_3$, and $\it{c}$-C$_3$H$_2$. The vertical line marks
5.9 km s$^{-1}$.}
\label{fg:H2C6}
\end{figure}

\begin{figure}[tbp]
\caption{The antenna temperatures for the ortho levels of H$_2$C$_6$ (solid
lines) and para levels of H$_2$C$_4$ (dashed lines) calculated using a LVG
excitation model and approximating the energy levels with a simple linear rotor
(see text) as a function of $J$.  The parameters are T$_{kin} = 10$~K,
$n$(H$_2) = 10^4$ cm$^{-3}$, and $\Delta$V$/$$\Delta$L = 1 km s$^{-1}$
pc$^{-1}$.  Curves for three fractional abundances (in units of km s$^{-1}$
pc$^{-1}$) are shown for each molecule.  The observed T$_A$ for H$_2$C$_4$
(filled triangle) and H$_2$C$_6$ (filled circles) are marked in the figures
along with their $1 \sigma$ error bars.  Total fractional abundances need to
include a factor for the ortho and para fractions, the velocity gradient, and
antenna efficiency (see text).}
\label{fg:H2CnExcitation}
\end{figure}

\begin{figure}[tbp]
\caption{The fractional abundances of five carbon chains, C$_{2n}$H,
C$_{2n+1}$H, H$_{2}$C$_{2n+1}$, H$_{2}$C$_{2n}$, and HC$_{2n+1}$N as a function
of number of carbons C$_n$.  These are taken from the early time solutions for
the gas phase chemical model of Millar et al.  for a cold cloud core with
T$_{kin}$ = 10K and n(H$_2$)=10$^4$ cm$^{-3}$.	The observed fractional
abundances of H$_{2}$C$_{4}$ and H$_{2}$C$_{6}$, assuming $\Delta$V$/$$\Delta$L
= 5 km s$^{-1}$ pc$^{-1}$, are marked by filled triangles.}
\label{fg:H2CnChemistry}
\end{figure}

\end{document}